\documentclass[a4paper,prb,aps,superscriptaddress,showpacs,twocolumn]{revtex4}
\usepackage{amssymb}
\usepackage{amsmath}
\usepackage{graphicx}
\usepackage{hyperref,hypernat}

\begin{document}
\title{Josephson current through a molecular transistor in a dissipative environment}
\author{Tom\'{a}\v{s} Novotn\'{y}}
\email{novotny@fys.ku.dk}
\affiliation{Nano-Science Center, Niels
Bohr Institute, University of Copenhagen, Universitetsparken 5,
DK-2100~Copenhagen \O , Denmark} \affiliation{Department of
Electronic Structures, Faculty of Mathematics and Physics, Charles
University, Ke Karlovu 5, 121 16 Prague, Czech Republic}
\author{Alessandra Rossini}
\affiliation{Nano-Science Center, Niels Bohr Institute, University of Copenhagen,
Universitetsparken 5, DK-2100~Copenhagen \O , Denmark}
\affiliation{Dipartimento di Fisica, Universit\`{a} di Milano, via Celoria 16, 20133
Milano, Italy}
\author{Karsten Flensberg}
\affiliation{Nano-Science Center, Niels Bohr Institute, University of Copenhagen,
Universitetsparken 5, DK-2100~Copenhagen \O , Denmark}
\date{\today }

\begin{abstract}
We study the Josephson coupling between two superconductors
through a single correlated molecular level, including Coulomb
interaction on the level and coupling to a bosonic environment.
All calculations are done to the lowest, i.e., the fourth, order
in the tunneling coupling and we find a suppression of the
supercurrent due to the combined effect of the Coulomb interaction
and the coupling to environmental degrees of freedom. Both
analytic and numerical results are presented.
\end{abstract}

\pacs{74.50.+r, 85.25.Cp, 73.40.Gk, 74.78.Na}
\maketitle

\section{Introduction}

Mesoscopic systems connected to superconducting leads have been
investigated for a number of years. If the mesoscopic system
itself is superconducting the transport is influenced by the
so-called parity effect in the Coulomb blockade
regime.\cite{ave-prl-92,tuo-prl-92} For normal metal grains, the
effect of the superconductivity is merely to modify the tunneling
density of states due to the density of the superconducting leads,
thus introducing an additional gap in the current-voltage
characteristics.

Naturally, the Josephson current is also affected by the Coulomb
correlations, because the transfer of a Cooper pair charges the
mesoscopic system by two electron charges. This process is similar
to a cotunneling event, and in the limit of large Coulomb
repulsion, the path that involves double occupancy of the central
region is not allowed, which results in a suppression of the
Josephson coupling. This has been studied extensively in a number
of theoretical works, starting with the work of Shiba and
Soda\cite{shi-ptp-69} and Glazman and Matveev,\cite{glaz-jetpl-89}
who calculated the Josephson current in the limit of infinite
Coulomb repulsion and in perturbation theory, the leading order
being the fourth order in the tunneling amplitudes. Interestingly,
the Josephson current was shown to change its sign when the level
which supports the tunneling current becomes occupied. This
so-called $\pi$ junction behavior, which is consequently also
relevant in the Kondo regime, has been studied in a number of
papers.\cite{spi-prb-91,and-prb-95,cle-prb-00,roz-prb-01,vec-prb-03,sia-prl-04,zai-jltp-04}
For larger dots the interplay of the above mentioned parity effect
and the Josephson effect has also been addressed.\cite{bau-prb-94}

Experimentally, there has been a number of studies of metallic
wires connected to superconductors,\cite{sch-prl-97,sch-nat-98}
but only a few studies in the Coulomb blockade regime. Buitelaar
{\em et al.}\cite{buit-prl-02,buit-prl-03} have observed multiple
Andreev reflections and Kondo physics in carbon nanotube quantum
dots. The supercurrent was not directly observed.

In this paper, we study the Josephson coupling through a single
level system coupled to vibrational or dissipative environments.
This is relevant for molecular transistor systems where strong
influence of the coupling to various vibrational modes have been
observed.\cite{park-nat-00,yu-nl-04,yu-prl-04} In the case of normal
metal leads this has lead to a number of related theoretical
works.\cite{boe-epl-01,mcc-prb-03,fle-prb-03,braig-prb-03,bra-prb-04,mit-prb-04}
However, so far the combination of the vibrational coupling and
correlation transport has not been studied in the context of
supercurrent. Being a groundstate property the dissipative
environment is expected to have a more dramatic effect on the
supercurrent as compared to usual electron tunneling. This is indeed
what we find in the case of strong coupling, where the Franck-Condon
factors strongly suppress the supercurrent.

The paper is organized as follows. In the next section our model
Hamiltonian is presented and a convenient unitary transformation
is performed. In Sec.~III, the basic formula for the Josephson
current is derived to the fourth order in the tunneling
amplitudes, and the different ingredients in this formula are
discussed. Section IV gives the supercurrent without coupling to
the bosons, both with and without Coulomb interaction, while the
Sec.~V studies the full case and discusses different limiting
cases. Finally, in Sec.~VI our conclusions are stated. The
technical details of analytic calculations are put into two
Appendixes.

\section{Model Hamiltonian}

\begin{figure}[tbp]
\centering
\includegraphics[width=0.4\textwidth]{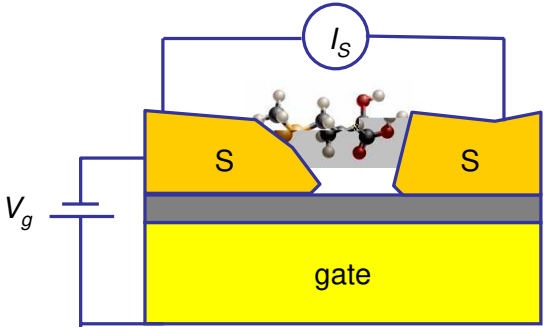}
\caption{Schematic picture of the physical setup described by the
model. The central region (molecule/quantum dot) is coupled to two
superconducting leads and can be independently gated by a gate.
Electrons entering the central region experience mutual Coulomb
interaction and interact with oscillatory modes --- either one
distinguished mode such as center-of-mass vibration of the
molecule as a whole or many modes due to intramolecular vibrations
or substrate phonons acting as a heat bath.} \label{fig0:model}
\end{figure}

The model we study in this paper consists of two superconducting
leads and a central region described by a single level with
Coulomb interaction and its charge occupation coupled to one or
many harmonic oscillator modes, see Fig.~\ref{fig0:model}.

The model Hamiltonian is given by
\begin{align}
H & =H_{0}+H_{T},
\end{align} where the unperturbed part of the
Hamiltonian is
\begin{align}
H_{0}  =H_{L}+H_{R}+H_{M},
\end{align}
with $H_{L,R}$ being the BCS Hamiltonians for the left and right
leads, respectively,
\begin{align}
H_{\alpha} &
=\sum_{k,\sigma=\uparrow,\downarrow}\varepsilon_{k\alpha}c_{k\alpha\sigma}^{
\dag}c_{k\alpha\sigma}-\left(\sum_{k}\Delta_{\alpha}c_{k\alpha\uparrow}^{
\dag}c_{-k\alpha\downarrow}^{\dag}+\mathrm{H.c.}\right),
\end{align}
with $\alpha=L,R$, and  the complex gap functions defined as
$\Delta_{\alpha} =|\Delta_{\alpha}|e^{i\phi_{\alpha}}$.
Furthermore, $H_M$ describes the central region
\begin{equation}\label{HM}
H_{M} =H_{M0}+H_\mathrm{vib}+H_{e-\mathrm{vib}},
\end{equation}
where $H_{M0}$ is the Hamiltonian for the electronic degrees of
freedom
\begin{equation}\label{HM0}
    H_{M0}=\xi\sum_{\sigma=\uparrow,\downarrow}d_{\sigma}^{\dag}d_{\sigma}+
Un_{\uparrow}n_{\downarrow},
\end{equation}
with $d_{\sigma},d_{\sigma}^\dagger$ being the operators for the
local level, and $n_\uparrow$ and $n_\downarrow$ the corresponding
spin-dependent occupations. We assume that the single-particle
energy level $\xi$ can be experimentally tuned by the gate, i.e.,
$\xi\equiv\xi(V_g)$, and present the results for the critical
Josephson current as functions of $\xi$. We have here defined the
origin (zero) of $\xi$ to be at the Fermi level of the
superconducting electrodes.  $H_\mathrm{vib}$ is the Hamiltonian
for the vibrational degrees of freedom
\begin{equation}\label{Hvib}
   H_\mathrm{vib}=\sum_\nu\left(\frac{p^2}{2m_\nu^{{}}}
   +\frac12 m_\nu^{{}}\omega_\nu^2x_\nu^2\right)=\sum_\nu\omega_\nu
   \left(a^\dagger_\nu
   a_{\nu}^{{}}+\frac12\right),
\end{equation}
and, finally, $H_{e-\mathrm{vib}}$ describes the coupling between
electronic and vibrational degrees of freedom. This coupling is
solely through the charge on the central system, i.e.,
\begin{equation}\label{Hevib}
H_{e-\mathrm{vib}}=(n_\uparrow +n_\downarrow)X,
\end{equation}
where
\begin{equation}\label{Xdef}
    X=\sum_\nu \lambda_\nu x_\nu.
\end{equation} In this model we neglect any modification of the tunneling
due to the vibrations (this is very often the relevant situation
and the model could easily be generalized to incorporate such a
dependence), and the tunneling Hamiltonian is therefore given by
\begin{equation}
H_{T}=\sum_{\alpha=L,R}(H_{T\alpha}^{+}+H_{T\alpha}^{-}),
\end{equation}
where $H_{T\alpha}^{+}=\left(H_{T\alpha}^{-}\right)^{\dagger}$ and
\begin{equation}
H_{T\alpha}^{-}=\sum_{\sigma=\uparrow,\downarrow}H_{T\alpha\sigma}^{-},\quad
H_{T\alpha\sigma}^{-}=\sum_{k}t_{k\alpha}c_{k\alpha\sigma}^{\dagger}d_{\sigma}.
\label{HTMdef}
\end{equation}

In the next section we will calculate the Josephson coupling using
perturbation theory in the tunneling. For this purpose, it is
convenient first to use a polaron representation, which transforms
the coupling term $H_{e-\mathrm{vib}}$ to a displacement operator
in the tunneling term.\cite{mahan,braig-prb-03}

The unitary transformation
\begin{equation}
\tilde{H}=SHS^{\dagger},\quad
S=e^{-iA(n_\downarrow+n_\uparrow)^{{}}},
\end{equation}
where
\begin{equation}\label{Adef}
A=\sum_\nu \ell_\nu p_\nu,\quad
\ell_\nu=\frac{\lambda_\nu}{m_\nu\omega_\nu^2},
\end{equation}
removes the coupling term $H_{e-\mathrm{vib}}$ from the
Hamiltonian at the expense that the tunneling term acquires an
oscillator displacement operator, so that Eq.~\eqref{HTMdef}
becomes
\begin{equation}
\label{HTMdefny}
\tilde{H}_{T\alpha\sigma}^{-}=SH_{T\alpha\sigma}^{-}S^\dagger
=\sum_{k}t_{k\alpha}c_{k\alpha\sigma}^{\dagger}d_{\sigma}e^{iA}.
\end{equation}
Furthermore, the transformation renormalizes the on-site energy
and the Coulomb interaction according to
\begin{equation}\label{uerenor}
    \tilde{\xi}=\xi-\frac12\sum_\nu \lambda_\nu\ell_\nu,
\quad \tilde{U}=U-\sum_\nu \lambda_\nu\ell_\nu.
\end{equation}
In the following we will skip the tildes and use just $\xi, U$
again but we mean the renormalized quantities. Using this
transformed Hamiltonian, we calculate the Josephson current to the
lowest order in the tunneling Hamiltonian in the following
sections.

\section{Josephson current}

The current operator for the current through contact $\alpha=L,R$
is $\dot{N}_\alpha$ (we use $\hbar=e=1$ throughout the whole
paper), where $N_\alpha$ is the operator of the number of
electrons in lead alpha. After the unitary polaron transformation
introduced previously, we hence obtain the current, $I_{\alpha}$,
as
\begin{equation}
I_{\alpha} =
i\left\langle\left[\tilde{H},N_{\alpha}\right]\right\rangle =
i\left\langle
\tilde{H}_{T\alpha}^{+}-\tilde{H}_{T\alpha}^{-}\right\rangle
=2\,\mathrm{Im}\,\left\langle \tilde{H}_{T\alpha}^{-}\right\rangle
.
\end{equation}
Performing the standard thermodynamic perturbation
expansion\cite{mahan,karsten} in the tunneling we obtain for the
Josephson current in the lowest non-vanishing order, which is the
fourth order, in $\tilde{H}_{T}$
\begin{equation}
\begin{split}
I_{\alpha}&=-2\,\mathrm{Im}\,\frac{1}{3!}\int_{0}^{\beta}d\tau_{1}\int_{0}^{
\beta}d\tau_{2}\int_{0}^{\beta}d\tau_{3}\\
&\quad\times\left\langle
T_{\tau}\left(\tilde{H}_{T}(\tau_{1})\tilde{H}_{T}(\tau_{2})\tilde{H}_{T}(\tau_{3})\tilde{H}_{T\alpha}^{-}\right)
\right\rangle_{0}.
\end{split}
\end{equation}
The Josephson current must involve two $\tilde{H}_{T}^{+}$ and two
$\tilde{H}_{T}^{-}$, which can be chosen in three ways, and hence
\begin{equation}
\begin{split}
I_{\alpha}&=-\mathrm{Im}\,\int_{0}^{\beta}d\tau_{1}\int_{0}^{
\beta}d\tau_{2}\int_{0}^{\beta}d\tau_{3}\\
&\quad\times\left\langle T_{\tau}\left(\tilde{H}_{T\bar{\alpha
}}^{+}(\tau_{1})\tilde{H}_{T\bar{\alpha}}^{+}(\tau_{2})\tilde{H}_{T\alpha}^{-}(\tau_{3})\tilde{H}_{T
\alpha}^{-}\right)\right\rangle_{0},
\end{split}
\end{equation}
where we also used that in order to have Cooper pair tunneling,
the $\tilde{H}_{T}^{+}$ must belong to the junction opposite to
where the Josephson current is ``measured" via
$\tilde{H}_{T\alpha}^-$, i.e., $\bar{\alpha}$ means the lead
opposite to $\alpha$. Because of spin symmetry we can choose the
spin of the last $\tilde{H}_T^-$ as, say, spin up, which then
means that the other $\tilde{H}_T^-$ carries spin down. In the
same way, the spin of the two $\tilde{H}_T^+$ can be chosen in two
ways. All in all, we thus obtain
\begin{equation}\label{total}
\begin{split}
I_{\alpha}
&=-4\,\mathrm{Im}\,\int_{0}^{\beta}d\tau_{1}\int_{0}^{\beta}d\tau_{2}\int_{0}^{\beta}d\tau_{3}
\\&\quad\times\left\langle T_{\tau}\left(\tilde{H}_{T\bar{\alpha}\downarrow}^{+}(\tau_{1})
\tilde{H}_{T\bar{\alpha}\uparrow}^{+}(\tau_{2})\tilde{H}_{T\alpha
\downarrow}^{-}(\tau_{3})\tilde{H}_{T\alpha\uparrow}^{-}\right)\right\rangle_{0}\\
&=-4\,\mathrm{Im}\,\sum_{k}\sum_{p}t_{p\bar{\alpha}}^{\ast}t_{-p\bar{\alpha}}^{\ast}
t_{-k\alpha}t_{k\alpha}\int_{0}^{\beta}d\tau_{1}\int_{0}^{\beta}d\tau_{2}\int_{0}^{\beta}d\tau_{3}
\\& \quad\times
\mathcal{F}_{k\alpha}(\tau_3)\mathcal{F}_{p\bar{\alpha}}^\ast(\tau_1-\tau_2)
\mathcal{B}(\tau_{1},\tau_{2},\tau_{3})\,\mathcal{D}(\tau_{1},\tau_{2},\tau_{3}),
\end{split}
\end{equation}
where $\mathcal{F}$ are the anomalous Green's functions of the
leads
\begin{equation}
    \mathcal{F}_{k\alpha}(\tau)
    =-\left\langle
    T_{\tau}\left(c^\dagger_{-k\alpha\downarrow}(\tau)c^\dagger_{k\alpha\uparrow}(0)\right)\right\rangle_0
\end{equation}
and where we define the following two functions pertaining to the
central region:
\begin{equation}\label{Bdef}
\mathcal{B}(\tau_{1},\tau_{2},\tau_{3})=\left\langle
T_{\tau}\left(d_{\downarrow}^{\dagger}(\tau_{1})d_{\uparrow}^{\dagger}(\tau_{2})
d_{\downarrow}(\tau_{3})d_{\uparrow}(0)\right)\right\rangle_{0}
\end{equation}
and
\begin{equation}\label{Ddef}
\mathcal{D}(\tau_{1},\tau_{2},\tau_{3})=\left\langle
T_{\tau}\left(e^{-iA(\tau_{1})}e^{-iA(\tau_{2})}e^{iA(\tau_{3})}
e^{iA}\right)\right\rangle _{0}.
\end{equation}
The first function $\mathcal{B}$ describes the propagation of a
Cooper pair through the central region, while the other function
$\mathcal{D}$ accounts for the corresponding shifts of the
oscillator degrees of freedom when the charge on the central
region is changed.

The anomalous Green's functions $\mathcal{F}$ are easily
calculated in the standard way,\cite{mahan,karsten} and we have
\begin{equation}\label{Fdef}
\mathcal{F}_{k\alpha}(\tau)=\frac{\Delta_{\alpha}^*}{2E_{k\alpha}}f_{\alpha}(E_{k\alpha},\tau),
\end{equation}
where
\begin{equation}\label{fdef}
     f_{\alpha}(E_{k\alpha},\tau)=e^{-E_{k\alpha}|\tau|}-2\cosh%
\left(E_{k\alpha}\tau\right)n_{F}(E_{k\alpha})
\end{equation}
and as usual
$E_{k\alpha}=\sqrt{\varepsilon_{k\alpha}^{2}+|\Delta_{\alpha}|^{2}}$.
Throughout, we will assume low temperatures such that
$|\Delta_{L,R}|\beta\gg1$, and we can thus approximate
\begin{equation}\label{fapprox}
f_{\alpha}(E_{k\alpha},\tau)\approx
e^{-E_{k\alpha}|\tau|}-e^{-E_{k\alpha}(\beta-|\tau|)}.
\end{equation}

In expression \eqref{total} for the current there are two sums
over states in the superconductors, which define the tunneling
density of states (TDOS) as
\begin{equation}\label{Jdos}
\Gamma_{\alpha}(\varepsilon)\equiv
2\pi\sum_{k}t_{k\alpha}t_{-k\alpha}\delta(\varepsilon-\varepsilon_{k\alpha})
=2\pi\sum_{k}|t_{k\alpha}|^2\delta(\varepsilon-\varepsilon_{k\alpha}).
\end{equation}
For a small central region, where the coupling is point-like, we
can approximate $t_k$ by a constant, which gives a weak energy
dependence of $\Gamma$.

\subsection{Critical current}

The Josephson current is a function of the phase-difference
between the two superconductors and using Eqs.~\eqref{total},
\eqref{Fdef}, and \eqref{Jdos}, we have
\begin{equation}\label{IIc}
    I_\alpha=I_c \sin\theta,
\end{equation}
where the phase difference
$\theta=\phi_{\bar{\alpha}}-\phi_\alpha$. Finally, the critical
current $I_c$ is given by
\begin{equation}\label{Icdef}
\begin{split}
I_c&=-\frac{1}{\pi^2}\int_{0}^{\beta}d\tau_{1}\int_{0}^{\beta}d\tau_{2}\int_{0}^{\beta}d\tau_{3}
\int_{-\infty}^{\infty} d\varepsilon \int_{-\infty}^{\infty} d\varepsilon'\\
&\quad\times\Gamma_L(\varepsilon)\Gamma_R(\varepsilon')\frac{|\Delta_L\Delta_R|}{4EE'}
f_L(E,\tau_3) f_R(E',\tau_1-\tau_2)
\\&\quad\times\mathcal{B}(\tau_{1},\tau_{2},\tau_{3})\,\mathcal{D}(\tau_{1},\tau_{2},\tau_{3}).
\end{split}
\end{equation}
This expression forms the basis for the further calculations in
this paper. In fact, from now on we will assume that the two
tunneling densities of states are energy independent.

The value of $I_c$ may come out negative and we will see in the
following that it really does so due to the Coulomb interaction.
The case with $I_c<0$ is called a $\pi$ junction. This terminology
originates from an equivalent description of the Josephson
junction in terms of total energy (or free energy at nonzero
temperature) of the junction as a function of the phase difference
$E(\theta)$. Since $I_{\alpha}= 2dE(\theta)/d\theta$ the total
energy reads $E(\theta)=-(I_c/2)\cos\theta$ and reaches minimum at
$\theta=0$ for $I_c>0$ or $\theta=\pi$ for $I_c<0$, respectively
(i.e., the ground state of the junction corresponds to
equal/opposite phases in the two leads). The $\pi$ junction
behavior has been noted in a number of papers.
\cite{shi-ptp-69,glaz-jetpl-89,spi-prb-91,and-prb-95,cle-prb-00,roz-prb-01,vec-prb-03,sia-prl-04,zai-jltp-04}
The origin of this sign change is the blocking of channels for the
Cooper pair exchange when $U$ is large. See
Ref.~\onlinecite{spi-prb-91} for a detailed account.

Ideally the Josephson current can be measured in a current-bias
setup where there is no voltage drop across the junction until the
critical current is reached. However, in practice current-bias is
difficult to achieve for large resistance junctions such as these
single-electron devices and instead a voltage is swept across the
junction and the critical current must be determined as half of
the area of the (ideally $\delta$-function-like) peak in the
$dI/dV$curve around $V=0$, which then, of course, only yields the
absolute value of $I_c$.

\subsection{Function $\mathcal{B}$}

Since the central region has interactions, we cannot use Wick's
theorem, and we must evaluate the function $\mathcal{B}$ using the
many-body states, of which there are four:
$|0\rangle,|\!\!\uparrow\rangle,|\!\!\downarrow\rangle,$ and $|\!\!\uparrow\downarrow%
\rangle.$ In Eq.~\eqref{Bdef} only $|\!\!\uparrow\rangle$ and $|\!\!\uparrow\downarrow%
\rangle$ contribute to the trace, i.e.,
$\mathcal{B}=\mathcal{B}_{1}+\mathcal{B}_{2}$, where
\begin{subequations}
\begin{align}
\mathcal{B}_{1} & =P_{\uparrow}\left\langle
\uparrow\!\left|T_{\tau}\left(d_{\downarrow}^{\dagger}(\tau_{1})d_{\uparrow}
^{\dagger}(\tau_{2})
d_{\downarrow}(\tau_{3})d_{\uparrow}(0)\right)\right|\!\uparrow\right\rangle_0,\\
\mathcal{B}_{2} & =P_{\uparrow\downarrow}\left\langle
\uparrow\downarrow\!\left|T_{\tau}\left(d_{\downarrow}^{\dagger}(\tau_{1})
d_{\uparrow}^{\dagger}(\tau_{2})d_{\downarrow}(\tau_{3})d_{\uparrow}(0)
\right)\right|\!\uparrow\downarrow\right\rangle_0,
\end{align}
\end{subequations}
with
\begin{subequations}
\begin{align}\label{Pup}
P_{\uparrow}&=P_{\downarrow}=\frac{e^{-\beta\xi}}{1+2e^{-\beta\xi}+e^{-\beta
E_{2}}},\\\label{Pupdown} P_{\uparrow\downarrow}&=\frac{e^{-\beta
E_{2}}}{1+2e^{-\beta\xi}+e^{-\beta E_{2}}},
\end{align}
\end{subequations}
and $E_{2}=2\xi+U$. For $B_{1}$ only three orderings of the
operators give a nonzero result, and we find
\begin{equation}
\begin{split}
\mathcal{B}_{1} & =P_{\uparrow}\left\{
e^{\xi(\tau_{1}-\tau_{3}+\tau_{2})}\theta(\tau_{2}-\tau_{3})\theta(\tau_{3}-
\tau_{1})\right.\\
&\quad+e^{E_{2}(\tau_{2}-\tau_{3})}e^{\xi(\tau_{1}-\tau_{2}+\tau_{3})}\theta(%
\tau_{3}-\tau_{2})\theta(\tau_{2}-\tau_{1}) \\
&\quad+\left.e^{E_{2}(\tau_{1}-\tau_{3})}e^{\xi(\tau_{2}-\tau_{1}+%
\tau_{3})}\theta(\tau_{3}-\tau_{1})
\theta(\tau_{1}-\tau_{2})\right\}.
\end{split}
\end{equation}
Likewise, for $B_{2}$ we find
\begin{equation}
\begin{split}
\mathcal{B}_{2}&=-P_{\uparrow\downarrow}
\left\{e^{\xi(\tau_{2}-\tau_{3}-
\tau_{1})}e^{E_{2}\tau_{1}}\theta(\tau_{1}-\tau_{2})
\theta(\tau_{2}-\tau_{3})\right.\\
&\quad+e^{\xi(\tau_{1}-\tau_{2}-\tau_{3})}e^{E_{2}
\tau_{2}}\theta(\tau_{2}-\tau_{1})\theta(\tau_{1}-\tau_{3}) \\
&\quad
\left.+e^{(E_{2}-\xi)(\tau_{1}+\tau_{2}-%
\tau_{3})}\theta(\tau_{1}-\tau_{3})
\theta(\tau_{3}-\tau_{2})\right\}.
\end{split}
\end{equation}

\subsection{Function $\mathcal{D}$}

We also calculate the function $\mathcal{D}$ in Eq.~\eqref{Ddef}
involving the bosonic degrees of freedom. Since the unperturbed
Hamiltonian is quadratic in boson operators, this is evaluated
as\cite{mahan}
\begin{equation}  \label{funct_F}
\begin{split}
\mathcal{D}=\exp\Big(&h(\tau_{1}-\tau_{3})
+h(\tau_{1})+h(\tau_{2}-\tau_{3})+h(\tau_{2})\\
&-h(\tau_{1}-\tau_{2}) -h(\tau_{3})\Big),
\end{split}
\end{equation}
where the function $h$ is defined as
\begin{equation}
h(\tau)=\left\langle T_{\tau}\big(A(\tau)A(0)\big)\right\rangle_0
-\left\langle A^{2}\right\rangle_0.
\end{equation}
When inserting the formula \eqref{Adef} for the operator $A$  one
obtains
\begin{equation}\label{hdef}
\begin{split}
h(\tau)&=\int_0^{\infty}d\omega \frac{J(\omega)}{\omega^2}
\left(n_{B}( \omega)(e^{\omega|\tau|}-1)\right.\\
&\qquad\qquad\qquad\
\left.+(1+n_{B}(\omega))(e^{-\omega|\tau|}-1)\right),
\end{split}
\end{equation}
with the spectral function of the bath $J(\omega)$
\begin{equation}
J(\omega) = \sum_{\nu}
\frac{\lambda_{\nu}^2}{2m_{\nu}\omega_{\nu}}\,\delta(\omega-\omega_{\nu}),
\end{equation}
and the Bose function $n_B(\omega)$
\begin{equation}
n_B(\omega)=\frac{1}{\exp(\beta\omega)-1}.
\end{equation}

In this paper, we concentrate on two cases --- either a single
important vibrational mode or a bath of harmonic modes. For the
latter case, we consider mainly the situation where the dispersion
relation corresponds to the so-called Ohmic case, equivalent to a
frequency independent damping coefficient or, equally, spectral
function $J(\omega)$ linear in frequency.

\subsubsection{Single oscillator case}

The case of a single oscillator with frequency $\omega_0$, mass
$m$ and coupling constant $\lambda$ corresponds to $J_{\mathrm{
osc}}(\omega)=g\omega_0^2\delta(\omega-\omega_0)$, where the
dimensionless $g=\lambda^2/2m\omega_0^3$ as in
Ref.~\onlinecite{braig-prb-03}, and we have
\begin{equation}
h_{\mathrm{osc}}(\tau)=g\big[n_{0}(e^{\omega_0|\tau|}-1)+
(1+n_{0})(e^{-\omega_0|\tau|}-1)\big],
\end{equation}
with $ n_0=n_B(\omega_0)$.

\subsubsection{Ohmic environment case}

The other generic case that we consider is that of Ohmic heat bath
in which case we use the spectral function
\begin{equation}\label{Johm}
    J_{\mathrm{ohm}}(\omega)=g\omega\exp(-\omega/ \omega_c)
\end{equation}
parametrized by the dimensionless interaction constant $g$ and the
upper cutoff frequency $\omega_c$. In this important case it is
possible to calculate at $T=0$ the correlation function $h(\tau)$
from Eq.~\eqref{hdef} analytically yielding
\begin{equation}
h_{\mathrm{ohm}}(\tau)=-g\ln(1+\omega_c|\tau|)\ .
\end{equation}

\section{Josephson current without dissipation}

As a reference for the discussion on the influence of coupling to
a dissipative environment, we discuss the case without coupling to
the bosonic environment. This we do in three steps: first we set
$U=0$, then we look at the infinite $U$ case and finally we give
the expression for the general case.

\subsection{No Coulomb interaction $U=0$}
\label{subsec:U0}

This result is derived in Appendix \ref{app:U0}, and the critical
current is found to be
\begin{equation}\label{Ic0}
I_{c}(\xi)=\frac{\Gamma_L\Gamma_R}{2(1-(\xi/\Delta)^2)}\Big(\frac{%
\tanh(\beta\xi/2)}{\xi}-\frac{\tanh(\beta\Delta/2)}{\Delta}\Big).
\end{equation}
We note that this expression diverges when $\xi=0$ and $T=0$,
which however is regularized by higher order terms in $\Gamma$'s,
see discussion in Appendix \ref{app:U0}. Furthermore, we see that
the critical current is always positive, which means that the
negative critical currents, i.e., the $\pi$ junction behavior,
found below are a result of the correlations. In
Fig.~\ref{fig1_no_phonons} this noninteracting result is compared
with two interacting cases $U=\Delta$ and $U\to\infty$.

\subsection{Strong Coulomb interaction $U\to\infty$}

Let $U\to\infty$ such that the doubly occupied state is taken out.
Hence
\begin{equation}
\mathcal{B}_{1}=P_{\uparrow}e^{\xi(\tau_{1}-\tau_{3}+\tau_{2})}\theta(\tau_{2}-%
\tau_{3})\theta(\tau_{3}-\tau_{1})\,\,,  \label{B1}
\end{equation}
and when inserting this into formula for the critical current
\eqref{Icdef}, performing the imaginary time integrations using
the approximation \eqref{fapprox} and $\beta\Delta\gg 1$ (so that
$\exp(-\beta E_{k\alpha})\approx 0$), we
find\cite{glaz-jetpl-89,and-prb-95,roz-prb-01}
\begin{equation}\label{current}
I_{c}(\xi)=\frac{\Gamma_L\Gamma_R}{\pi^2} N(\xi)
\end{equation}
where
\begin{equation}\label{Ndef}
N(\xi)\!=-\!\!\int_{|\Delta_L|}^\infty
\frac{|\Delta_L|dE}{\sqrt{E^2-|\Delta_L|^2}}\int_{|\Delta_R|}^\infty
\frac{|\Delta_R|dE'}{\sqrt{E'^{2}-|\Delta_R|^2}}\,C(E,E')
\end{equation}
and
\begin{equation}
\begin{split}
C(E,E') &\!=\!\frac{2e^{-\beta\xi}}{1+2e^{-\beta\xi}}
\left(-\frac{e^{\beta\xi}}{(E+E')
\left(E+\xi\right)\left(E'+\xi\right)}\right.\\
&\left.\quad+\frac{1}{(E+E')\left(E-\xi\right)\left(E'-\xi\right)}\right).
\end{split}
\end{equation}
At $T=0,$ this reduces to
\begin{equation}
C(E,E)=\left\{
\begin{array}{cc}
-2\left[(E+E')(E+\xi)(E'+\xi)\right]^{-1} &
\mathrm{for}\,\xi>0, \\[2mm]
\left[(E+E')(E-\xi)(E'-\xi)\right]^{-1} & \mathrm{for}\,\xi<0.
\end{array}
\right.
\end{equation}
The integral in Eq.~\eqref{Ndef} can be performed analytically at
$T=0$, and $|\Delta_L|=|\Delta_R|=\Delta$ yielding
\begin{equation}\label{Nrel}
N(\xi)=\left\{
\begin{array}{cc}
2\,n(\xi/\Delta)/\Delta & \mathrm{for}\quad\xi>0, \\[2mm]
-n(-\xi/\Delta)/\Delta & \mathrm{for}\quad\xi<0.
\end{array}
\right.
\end{equation}
The dimensionless function $n(x)$ defined as ($x>-1$)
\begin{equation}\label{funct_n}
n(x)\!=\!\int_1^{\infty}\!\frac{du}{\sqrt{u^2-1}}\int_1^{\infty}\!\frac{dv}{\sqrt{v^2-1}}\frac{1}{(u+v)(u+x)(v+x)},
\end{equation}
can be expressed by (see Appendix \ref{app:Ufinite})
\begin{equation}  \label{funct_j}
 n(x) =\frac{\frac{\pi^{2}}{4}(1-x)-\arccos^2 x}{x(1-x^{2})},\quad\text{with }\,-1 < x,
\end{equation}
where the analytic continuations of $\arccos x= i\ln(x+\sqrt{x^2-1})$ for $%
x>1$ is understood. The function $n(x)$ is always positive, it
diverges at $x\to-1^+$, and then smoothly decays for increasing
$x$ with the asymptote $n(x)\sim \pi^2/4x^2$ for $x\to\infty$. The
expression \eqref{Nrel} is compared with the noninteracting and
finite $U$ results in Fig.~\ref{fig1_no_phonons}. We see that the
magnitude of the critical current is highly suppressed by the very
strong Coulomb interaction.

For finite but small temperatures $\beta\Delta\gtrsim 10$ so that
our approximations are still valid, we can write for the magnitude
of the Josephson current (for $|\xi|<\Delta$ only, otherwise the
zero temperature expression should be used)
\begin{equation}\label{temperature}
I_{c}(\xi)=\frac{2\Gamma_L\Gamma_R}{\pi^2\Delta}\frac{n(\xi/\Delta)-e^{-\beta\xi}n(-%
\xi/\Delta)}{(1+2e^{-\beta\xi})}.
\end{equation}
The finite temperature behavior in the limit $U\to\infty$ is
illustrated in more detail in Fig.~\ref{fig2_temp_dep} for three
values of temperature. In the lowest temperature curve
$\beta\Delta=10$ we also compare the analytic expression
\eqref{temperature} with a direct numerical evaluation of the
triple imaginary-time integral (Eq.~\eqref{current_eq} with
$\mathcal{D}\equiv 1$) routinely used for the dissipative cases.
We see an excellent agreement between the two methods.

In an experiment with a single Josephson junction the absolute
values of the presented curves would be measured. This would give
curves with a dip down to zero at $\xi=0$ (for a finite
temperature) and with asymmetric shoulders around the dip with the
ratio between the shoulder heights being 2. Even though the dip
may be smeared in the experiment for low enough temperatures the
asymmetric shoulders should persist thus revealing the crossover
to the $\pi$ junction regime.

\subsection{Finite Coulomb interaction $U$}

\begin{figure}[tbp]
\centering
\includegraphics[width=0.5\textwidth]{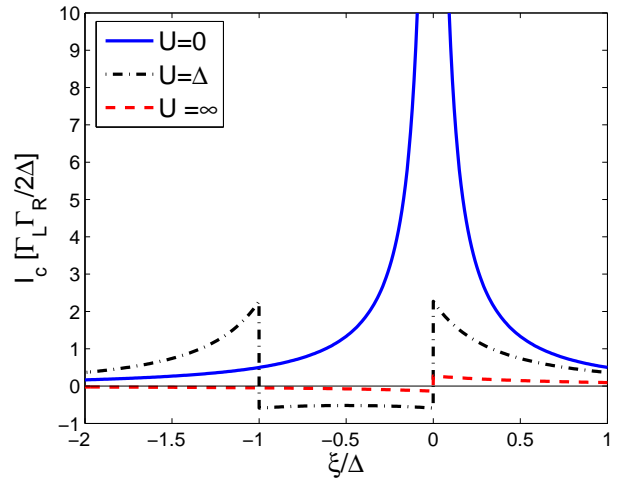}
\caption{(Color online) Josephson current dependence on the
Coulomb interaction strength at zero temperature. Shown are the
critical Josephson current through an Anderson level with no
Coulomb interaction $U=0$ (full line), with moderate interaction
$U=\Delta$ (dash-dotted line), and with infinite repulsion
$U\to\infty$ (dashes). The interacting cases exhibit the
phenomenon of the $\protect\pi$ junction for $-U<\protect\xi<0$,
but the overall magnitude of the current decreases largely with
the increasing interaction strength.} \label{fig1_no_phonons}
\end{figure}

\begin{figure}[tbp]
\centering
\includegraphics[width=0.5\textwidth]{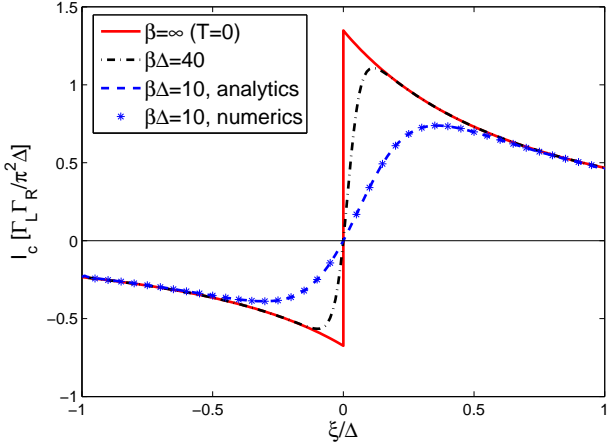}
\caption{(Color online) Josephson current dependence on the temperature for $%
U\to\infty$. We show the temperature dependence of the critical
Josephson current for $|\protect\xi|<\Delta$ for three different temperatures $T=0\ (%
\protect\beta\to\infty)$ (full line), $\protect\beta\Delta=40$
(dash-dotted line), and $\protect\beta\Delta=10$ (dashed line with
asterisks). In the last case which illustrates roughly the highest
temperature achievable within the approximations used the analytic
and numerical results are compared.} \label{fig2_temp_dep}
\end{figure}

For a finite value of $U$ we have to consider all terms of
$B=B_1+B_2$ in an analogous way as previously the first one in the
$U\to\infty$ case. Neglecting again terms $e^{-\beta E_{k,p}}$, we
recover the formula \eqref{current} for the current, but with the
function $C(E,E')$ in Eq.~\eqref{Ndef} replaced by
\begin{equation}\label{CC}
    C(E,E')=\frac{2}{Z(E+E')}L(E,E'),
\end{equation}
where $Z=\big(1+2e^{-\beta\xi}+e^{-\beta(2\xi+U)}\big)$ and
\begin{equation}\label{funct_L}
\begin{split}
L(E,E') &= -\frac{1}{(E+\xi)(E'+\xi)}
+\frac{e^{-\beta\xi}}{(E-\xi)(E'-\xi)}\\
&\quad+\frac{2\,e^{-\beta(2\xi+U)}(E+E')}{(2\xi+U)(\xi+U-E)(\xi+U-E')}\\
&\quad-\frac{2(E+E')}{(2\xi+U)(\xi+E)(\xi+E')}\\
&\quad+\frac{2e^{-\beta\xi}}{(E'-\xi)(E'+\xi+U)}\\
&\quad+\frac{2e^{-\beta\xi}}{(E-\xi)(E+\xi+U)}\\
&\quad+\frac{e^{-\beta\xi}}{(E+\xi+U)(E'+\xi+U)}\\
&\quad-\frac{e^{-\beta(2\xi+U)}}{(\xi+U-E)(\xi+U-E')}.
\end{split}
\end{equation}
The resulting integral does not seem to be analytically calculable
in the whole $\xi$-range, not even at $T=0$ and
$|\Delta_L|=|\Delta_R|=\Delta$. Yet, in that limit one can achieve
significant simplifications at least for some $\xi$'s which even
allow us to evaluate the $n(x)$ function defined by
Eq.~\eqref{funct_n} yielding Eq.~\eqref{funct_j}. The details of
those calculations can be found in Appendix \ref{app:Ufinite}. We
have, however, calculated the critical current numerically, and an
example is shown in Fig.~\ref{fig1_no_phonons} for $U=\Delta$.

\section{Josephson current with dissipation}

Next, we study how the critical Josephson current changes, when
coupling to vibrational modes is included. In order to do that, we
will perform a numerical integration of the three imaginary time
integrals. For this purpose we first find (again assuming
$\Gamma_\alpha(\varepsilon)$ constant)
\begin{equation}
\begin{split}
\mathcal{H}_\alpha(\tau)&\equiv
\int_{|\Delta_{\alpha}|}^{\infty}dE \frac{1}
{\sqrt{E^{2}-|\Delta_{\alpha}|^{2}}}\left[e^{-E|\tau|}-e^{-E(\beta-|\tau|)}\right] \\
& =K_{0}(|\tau\Delta_{\alpha}|)-K_{0}((\beta-|\tau|)
|\Delta_{\alpha}|),
\end{split}
\end{equation}
where $K_{0}(x)$ is the modified Bessel function of the second kind.
The expression for the critical current thus reads
\begin{equation}\label{current_eq}
\begin{split}
I_{c}&=-\frac{\Gamma_L\Gamma_R|\Delta_L\Delta_R|}{\pi^{2}}\int_{0}^{\beta}d\tau_{1}\int_{0}^{
\beta}d\tau_{2}\int_{0}^{\beta}d\tau_{3}\,\mathcal{H}_{L}(\tau_{1}-\tau_{2})\\
&\quad\times\mathcal{H}_{R}(\tau_{3})\mathcal{B}(\tau_{1},\tau_{2},\tau_{3})\mathcal{D}(\tau_{1},\tau_{2},\tau_{3}).
\end{split}
\end{equation}

We evaluated $I_c$ numerically for a number of different cases
with the qualitatively same results showing that the coupling to
oscillator mode(s) suppresses the magnitude of the Josephson
current. There is no apparent difference between the single mode
and Ohmic heat bath case which is in a clear contrast with
nonequilibrium transport studies where the character of the phonon
spectrum plays a crucial role in the current-voltage
characteristics. Below, just for simplicity, we only present
results for the symmetric case $|\Delta_L|=|\Delta_R|=\Delta$ at
zero temperature $T=0$ and for infinite Coulomb interaction
$U\to\infty$.

\subsection{Low-frequency phonons}

\begin{figure}[tbp]
\centering
\includegraphics[width=0.5\textwidth]{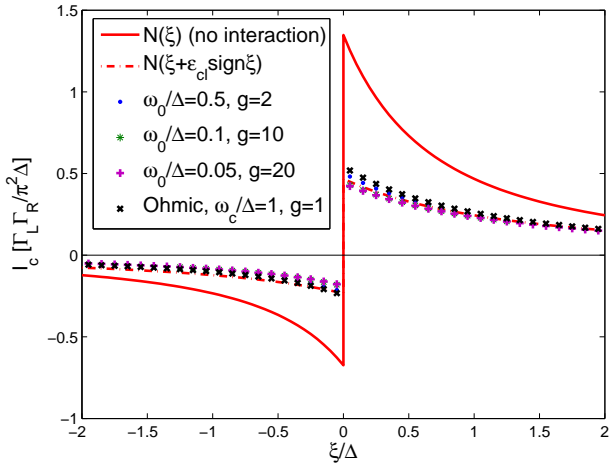}
\caption{(Color online) Josephson current for low-frequency phonons at $%
T=0,\,U\to\infty$. As long as the spectrum of the phonon mode(s)
is well below the superconducting gap, i.e.,
$\protect\omega_{0,c}\ll\Delta$, the
critical Josephson current only depends on the integral $\protect\varepsilon_{\rm cl}=%
\protect\int_0^{\infty}d\protect\omega
J(\protect\omega)/\omega=g\protect\omega_{0,c}$ via the shifted
dissipationless function $N(\protect\xi+\protect\varepsilon_{\rm
cl}\,\mathrm{sgn}\protect\xi)$ (dash-dotted line).
Numerical results are shown for a single phonon mode with $\protect\omega%
_0=0.5\Delta,\,g=2$ (dots), $\protect\omega_0=0.1\Delta,\,g=10$ (asterisks),
$\protect\omega_0=0.05\Delta,\,g=20$ (pluses), and for the Ohmic bath with $%
\protect\omega_c=\Delta,\,g=1$ (crosses). All of these cases yield basically
the same result very well captured by the analytic expression.}
\label{fig3_low_freq}
\end{figure}

If the spectrum of the oscillator mode(s) is well below the
superconducting gap $\omega_{0,c}\ll\Delta$ we can find an
approximate analytic expression for the $I_c(\xi)$ with the help
of the function $n(\xi)$ of the $U\to\infty$ case with no phonons,
Eq.~\eqref{funct_j}. To this end we study the current formula
\eqref{current_eq} when we plug into it the expressions for $
\mathcal{B}(\tau_1,\tau_2,\tau_3)$ \eqref{B1} and
$\mathcal{D}(\tau_1,\tau_2,\tau_3)$ \eqref{funct_F} and consider
the above limit. For $\xi\leq 0$ we notice that the step functions
of Eq.~\eqref{B1} and the fast decaying functions
$\mathcal{H}_{\alpha}(\tau)$ in Eq.~\eqref{current_eq} limit the
relevant contributions to the three-dimensional integral to values
of $ \tau_1,\tau_2,\tau_3$ small compared to
$\min\{1/|\xi|,1/\Delta\}$. For that reason one can perform the
Taylor expansion in Eq.~\eqref{funct_F} of the $h(\tau)$ functions
given by Eq.~\eqref{hdef} which gives
\begin{equation}  \label{expansion}
h(\tau)\approx-|\tau|\,\int_0^{\infty}d\omega\, \frac{
J(\omega)}{\omega}\equiv-\varepsilon_{\rm cl} |\tau| ,
\end{equation}
where we have defined the quantity
\begin{equation}\label{Eddef}
    \varepsilon_{\rm cl}=\int_0^{\infty}d\omega\,\frac{
J(\omega)}{\omega}=\sum_{\nu}\frac{\lambda_{\nu}^2}{2m_{\nu}\omega_{\nu}^2}
=\sum_{\nu}\frac 12 m_{\nu}\omega_{\nu}^2l_{\nu}^2,
\end{equation}
being the classical displacement energy when the oscillators are
displaced by $l_{\nu}$ due to the force generated by a single
excess electron. Putting the expansion into Eqs.~\eqref{funct_F}
and \eqref{current_eq} we get
\begin{equation}
\begin{split}
I_{c}(\xi)&=-\frac{\Gamma_L\Gamma_R\Delta^2}{\pi^{2}}\frac{e^{-\beta\xi}}{1+2e^{-\beta\xi}}
\int_{0}^{\beta}d\tau_{2}\int_{0}^{\tau_2}d\tau_{3} \int_{0}^{\tau_3}d\tau_{1}\\
&\quad\times
\mathcal{H}_{L}(\tau_{1}-\tau_{2})\mathcal{H}_{R}(\tau_{3})
\,e^{(\xi-\varepsilon_{\rm cl})(\tau_{1}-\tau_{3}+\tau_{2})},
\end{split}
\end{equation}
valid for $\text{ for } \xi\leq 0$. This in fact corresponds to
the dissipationless case with the replacement
$\xi\to\xi-\varepsilon_{\rm cl}$. In particular, in the zero
temperature limit we get $I_{c}(\xi)=-(\Gamma_L\Gamma_R/
\pi^{2}\Delta)n[(\xi-\varepsilon_{\rm cl})/\Delta]$ for $\xi\leq
0$. For $\xi\geq 0$ we have to combine the vanishing prefactor
$\exp(-\beta\xi)$ with the divergent integral (for
$\beta\to\infty$) to get a finite result. This can be done by the
substitutions $\tau_i^{\prime}=\beta-\tau_i,\ i=1,2,3$ which make
the integrand relevant only for small values of
$\tau_i^{\prime}$'s and, analogously to the previous derivation,
one finally finds in the zero temperature limit, that
$I_{c}(\xi)=(2\Gamma_L\Gamma_R/\pi^{2}\Delta)n[(\xi+\varepsilon_{\rm
cl})/\Delta]$ valid for $\xi\geq 0$. In total, the critical
Josephson current in case of coupling to low-frequency oscillator
modes can be expressed as (remember the relation \eqref{Nrel}
between $N(\xi)$ and $n(x)$)
\begin{equation}\label{Iclow}
I_c(\xi)=\frac{\Gamma_L\Gamma_R}{\pi^{2}\Delta}N(\xi+\varepsilon_{\rm
cl}\,\mathrm{sgn}\xi).
\end{equation}
This result is illustrated in detail in Fig.~\ref{fig3_low_freq}.

We point out that even in case of finite temperature there is no
residual temperature dependence due to coupling to the bath even
though the low-frequency modes could be significantly populated.
This is seen from the
expansion \eqref{expansion} where the thermal occupation factors $%
n_B(\omega)$ cancelled out. Thus, the only temperature dependence
would be the one stemming from the occupation factors exactly as
in the dissipationless case, Eq.~\eqref{temperature} and
Fig.~\ref{fig2_temp_dep}.

\subsection{High-frequency single oscillator mode}

\begin{figure}[tbp]
\centering
\includegraphics[width=0.5\textwidth]{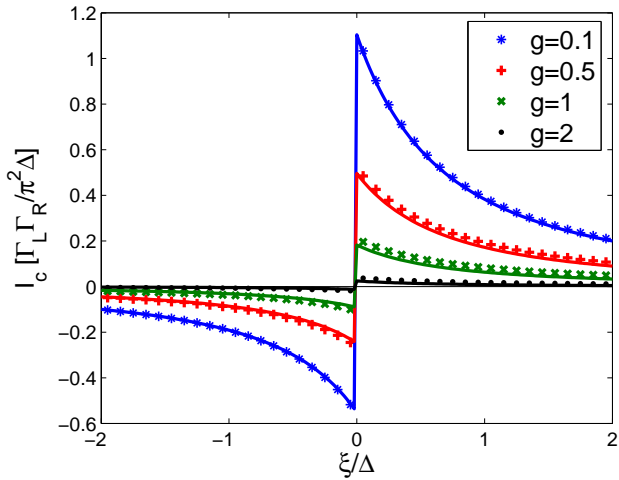}
\caption{(Color online) Josephson current for high-frequency
single oscillator at $T=0,\,U\to\infty,
\protect\omega_0=20\Delta$. For large enough
$\protect\omega_0\gg\Delta,\protect\xi$ the Josephson current is
expected to be suppressed just by the factor $\exp(-2g)$ compared
to the case without the phonon. This is shown in the figure where
the numerical results for several values of the coupling constant
$g=0.1$ (asterisks), $g=0.5$ (pluses), $g=1$ (crosses), and $g=2$
(dots) are compared with the analytic expressions given by the
corresponding lines.} \label{fig4_high_freq}
\end{figure}
For the high-frequency single phonon mode $\omega_0\gg\Delta,\xi$
we expect suppression of the critical Josephson current due to the
fact that only transport through the ground oscillator state is
allowed since the (virtual) involvement of the excited states
would be further suppressed by a factor $\Delta/n\omega_0\ll 1\
(n=1,2,3,\dots)$. The transport through the ground state is then
diminished by the overlap factors $f_{00}=\langle 0|e^{-i l_0
p_0}|0\rangle_0$, more precisely the Josephson current should be
suppressed by a factor of order
\begin{equation}\label{f004}
|f_{00}|^4=\left[\exp\left(
-\frac12\int_0^\infty\frac{d\omega}{\omega^2}J(\omega)\right)\right]^4=e^{-2g}.
\end{equation}
At $T=0$ and for $\omega_0\gg\Delta,\xi$ the function
$h_{\mathrm{osc}}(\tau)=g(e^{-\omega_0|\tau|}-1)$ changes very
fast for small $\tau$'s of the order $\tau\sim 1/\omega_0$ which
is irrelevant for the integral \eqref{current_eq}. For larger
$\tau$'s the function $h_{\mathrm{osc}}(\tau)\sim -g$ is constant
and thus we get from Eq.~\eqref{funct_F} that the total effect of
the phonon on the Josephson current is just a constant factor of
$e^{-2g}$ multiplying the dissipationless case. This effect is
shown in Fig.~\ref{fig4_high_freq} for different values of the
coupling constant $g$ at fixed $\omega_0=20\Delta$. The results do
not depend on the value of $\omega_0$ provided it is high enough,
which depends on the value of $g$, as expected (not shown). One
can see that the approximation gets worse with increased $g$ (for
fixed $\omega_0$) which can be explained by the increased
contribution of higher order virtual processes favored by the
larger value of the coupling constant.

\subsection{Ohmic bath with high-frequency cutoff}

\begin{figure}[tbp]
\centering
\includegraphics[width=0.5\textwidth]{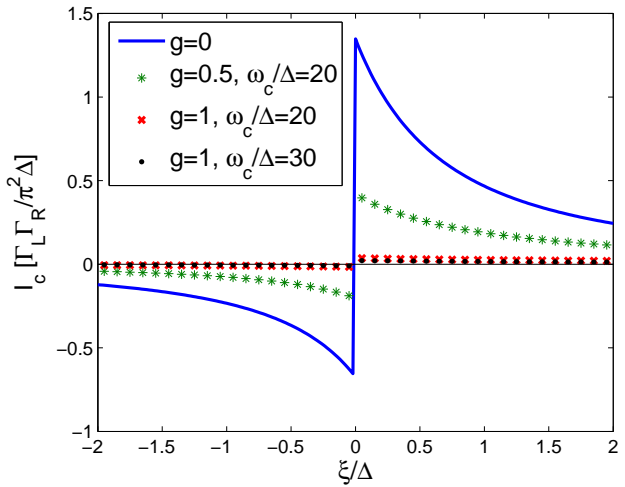}
\caption{(Color online) Josephson current for
high-frequency-cutoff Ohmic bath at $T=0,\,U\to\infty$ and for
several values of the coupling constant $g$ and cutoff frequency
$\omega_c$. We see a significant suppression of the critical
current already for intermediate coupling $g=1$.}
\label{fig5_high_freq_ohm}
\end{figure}

The case of Ohmic bath with large cutoff energy may be the
physically most relevant one. Unfortunately, there is no simple
semianalytic theory for this case and thus we have to rely mainly
on the numerical results which are summarized in
Fig.~\ref{fig5_high_freq_ohm}. One should notice the very strong
dependence on the coupling constant $g$. Even for intermediate
coupling strength $g\sim 1$ we get a significant suppression of
$I_c$.

We can give a qualitative explanation of the strong suppression
due to the large frequency part of the phonon spectrum,
$\omega>\Delta$. Roughly, we capture the effect of these
high-frequency modes by a suppression factor similar to
Eq.~\eqref{f004}. This factor is estimated as
\begin{equation}
|\tilde{f}_{00}|^4\sim\left[\exp\left(
-\frac{g}{2}\int_{\Delta}^{\omega_c}\frac{d\omega}{\omega}\right)\right]^4
=\left(\frac{\Delta}{\omega_c}\right)^{2g}.
\end{equation}
Thus for large $\omega_c$ and/or $g$ the supercurrent is
suppressed quite severely, in qualitative agreement with the
results presented in Fig.~\ref{fig5_high_freq_ohm}.

\section{Summary and discussion}

We have calculated the Josephson current through a single
correlated level with coupling to external bosonic degrees of
freedom representing, e.g., a system consisting of a molecular
transistor with a number of internal vibrational modes and coupled
to the phonons of the substrate.

First, we have studied the case without coupling to vibrations.
This situation has been studied previously in a number of papers,
but we have derived new analytic formulae. The effect of the
Coulomb interaction is to strongly suppress the Josephson current
at the charge degeneracy point, and since this is where the
junction also crosses over to $\pi$ junction behavior, the
critical current in fact goes to zero at this point. One could
check this behavior using, e.g., nanotube devices coupled to
superconductors as in Refs.~\onlinecite{buit-prl-02},
\onlinecite{buit-prl-03} by tuning the gate voltage across the
charge degeneracy point. Also the temperature dependence predicted
here could be experimentally verified.

In the second part of the paper we have included the coupling to
environmental modes and discussed different limits. Coupling to
low-frequency phonons does not have a severe influence on the
Josephson coupling, since it only shifts the argument of the
dissipationless formula from the single particle energy by the
classical displacement energy. In contrast, a strong coupling to
high-energy phonons suppresses the supercurrent quite
substantially. This is because the oscillators are displaced twice
during the transfer of Cooper pair, and because the supercurrent
requires the final state of the oscillator to be identical to the
initial one, the transfer is suppressed by twice the exponential
Franck-Condon factor $e^{-2g}$. Therefore it might be difficult to
observe Josephson tunneling current in devices with a strong
electron-vibron coupling.

Throughout the paper we have used lowest-order  perturbation
theory in the tunneling coupling. For stronger tunneling coupling
one expects the correlation effect due to the vibrations to become
smaller, because the charge on the level is no longer well
defined. This means that a mean-field treatment becomes adequate,
when $\Gamma>\varepsilon_{\rm cl}$. How to describe this
transition theoretically is an interesting problem.

\begin{acknowledgments}
The work of T.~N.\ is a part of the research plan MSM 0021620834
that is financed by the Ministry of Education of the Czech
Republic, while A.~R.\ and K.~F.\ were partly supported by the EC
FP6 funding (Contract No.~FP6-2004-IST-003673,
CANEL).\footnote{This publication reflects the views of the
authors and not necessarily those of the EC. The community is not
liable for any use that may be made of the information contained
herein.}
\end{acknowledgments}

\appendix
\section{Noninteracting case $U=0$}
\label{app:U0}

The noninteracting case is exactly solvable for any values of
parameters.\cite{beenakker,lin-ctp-02} Here, we only give a brief
sketch of the solution and the summary of the results in the limit
of small $\Gamma$ relevant for our study. Since the model for
$U=0$ has a quadratic Hamiltonian, it can be solved by the
equation of motion technique for the Matsubara Green function. We
introduce an infinite vector $ \alpha=(d_{\uparrow},
d^{\dag}_{\downarrow},c_{kL\uparrow}^{{}},c^{\dag}_{-kL\downarrow},
c_{kR\uparrow}^{{}},c^{\dag}_{-kR\downarrow})^T$ generalizing the
standard Nambu formalism to the case of the dot plus two leads.
Defining the corresponding thermal Green function
$\mathcal{G}(\tau)=-\langle
T_{\tau}(\alpha(\tau)\alpha^{\dag})\rangle$ satisfying the
equation of motion
$\frac{d}{d\tau}\mathcal{G}(\tau)=-\delta(\tau)\mathrm{1}+\mathrm{M}
\cdot\mathcal{G}(\tau)$ with the matrix
\begin{equation}
\mathrm{M} =
\begin{pmatrix}
-\xi & 0 & -t_{kL}^* & 0 & -t_{kR}^* & 0 \\
0 & \xi & 0 & t_{-kL} & 0 & t_{-kR} \\
-t_{kL} & 0 & -\varepsilon_{kL} & \Delta_L & 0 & 0 \\
0 & t_{-kL}^* & \Delta_L^* & \varepsilon_{kL} & 0 & 0 \\
-t_{kR} & 0 & 0 & 0 & -\varepsilon_{kR} & \Delta_R \\
0 & t_{-kR}^* & 0 & 0 & \Delta_R^* & \varepsilon_{kR}
\end{pmatrix}
\end{equation}
we can express the Josephson current as
\begin{equation}
I_L = -4\,\mathrm{Im}\sum_k t_{-kL} \mathcal{G}_{42}(\tau\to 0^+)
\end{equation}
(factor of 2 for spin degeneracy). Going to the frequency picture ($\frac{d}{%
d\tau}\rightarrow -i\omega_n$) and using the partitioning scheme
\begin{equation}
\!\!\begin{pmatrix}
A & c \\
d & B
\end{pmatrix}
^{-1}\!\!\!\!=\!\!
\begin{pmatrix}
(A-cB^{-1}d)^{-1} & -A^{-1}c(B-dA^{-1}c)^{-1} \\
-B^{-1}d(A-cB^{-1}d)^{-1} & (B-dA^{-1}c)^{-1}
\end{pmatrix}
\end{equation}
together with the wide-band approximation
$\Gamma_{\alpha}=\text{const}$, and assuming the symmetric case
$\Gamma_L=\Gamma_R=\Gamma,\,\Delta_L=\Delta
e^{i\phi},\,\Delta_R=\Delta$ this set of linear equations gives
for the Josephson current
\begin{equation}\label{U_zero_sum}
I_L = \mathrm{Im}\frac{1}{\beta}\sum_{\omega_n}e^{i\omega_n 0^-}
\frac{
\Gamma^2\Delta^2e^{-i\phi}}{(\omega_n^2+\Delta^2)D(\omega_n)}
\end{equation}
with
\begin{equation}\label{Diwn}
D(\omega_n)=\omega_n^2\Big(1+\frac{\Gamma}{\sqrt{\omega_n^2+\Delta^2}}
\Big)^2+\xi^2+\frac{\Delta^2\Gamma^2\cos^2(\phi/2)}
{(\omega_n^2+\Delta^2)}.
\end{equation}
The zeros of $D(\omega_n)$ determine the Andreev bound states
discussed in Ref.~\onlinecite{vec-prb-03}. In the limit $\Gamma\to
0$ which we consider here the lowest order contribution to $I_L$
is proportional to $ \Gamma^2$ and can be obtained by setting
$\Gamma=0$ in $D(\omega_n)$. The sum can then be easily performed
yielding
\begin{equation}\
I_L=-\sin\phi\frac{\Gamma^2}{2(1-(\xi/\Delta)^2)}\Big(\frac{\tanh(\beta\xi/2)}{\xi}-\frac{\tanh(\beta\Delta/2)}{\Delta}\Big).
\end{equation}
For $k_BT\ll\Delta$ this expression is proportional to $\Gamma^2/k_BT$ at $%
\xi=0$. This would diverge in the $T=0$ limit. However, this
divergence is just an artifact of our perturbation theory and the
exact evaluation of the full expression \eqref{U_zero_sum} would
give a finite result even at $T=0,\,\xi\to 0$. The exact result is
essentially identical to our approximate one unless
$\xi,k_BT\lesssim\Gamma$ when the exact result gives
a saturation of the maximum Josephson current\cite{beenakker,lin-ctp-02} $%
I_c\approx\Gamma$ around $\xi=0$ for $\Gamma\ll\Delta$. For
nonsymmetric coupling to the leads, i.e., $\Gamma_L\neq\Gamma_R$
but still $|\Delta_L|=|\Delta_R|=\Delta$, we get within the
discussed lowest order approximation in $\Gamma$'s the same result
just with the replacement $\Gamma^2\to\Gamma_L\Gamma_R$ which is
used in the main text, Sec.~\ref{subsec:U0}.

\section{Finite $U$, evaluation of $n(x)$}
\label{app:Ufinite}

In this appendix we calculate the critical Josephson current for
finite value of the Coulomb interaction $U$ in the limit of zero
temperature $T=0$ and symmetric gap
$|\Delta_L|=|\Delta_R|=\Delta$. The critical current is given by
Eq.~\eqref{current} with the function $N(\xi)$ being determined in
this case by Eqs.~\eqref{Ndef}, \eqref{CC}, and \eqref{funct_L}
which further simplify in the considered limit of zero temperature
and symmetric gap so that we can write for $N(\xi)$:
\begin{widetext}
\begin{subequations}
\begin{align}
    N(\xi) &=\int_{\Delta}^{\infty} dE \int_{\Delta}^{\infty} dE' \frac{2\Delta^2}{\sqrt{E^{2}-\Delta^{2}}\sqrt{E'^{2}-\Delta^{2}}}
    \left[\frac{1}{(E+E') \left(E+\xi\right)\left(E'+\xi\right)}+\frac{2}{(2\xi+U)(\xi+E)(\xi+E')}\right]\nonumber\\
    &=2\left(\frac{n(\xi/\Delta)}{\Delta}+\frac{\arccos^2(\xi/\Delta)}{(\xi+U/2)(1-(\xi/\Delta)^2)}\right)
    \qquad\mathrm{for}\quad\xi>0,\label{finiteU}\\
    N(\xi) &=-\int_{\Delta}^{\infty} dE \int_{\Delta}^{\infty} dE' \frac{\Delta^2}{(E+E')\sqrt{E^{2}-\Delta^{2}}\sqrt{E'^{2}-\Delta^{2}}}
    \left[\frac{1}{\left(E-\xi\right)\left(E'-\xi\right)}+\frac{1}{(E+\xi+U)(E'+\xi+U)}\right.
    \nonumber\\
    &\quad +\left.\frac{4}{(E-\xi)(E+\xi+U)}\right]=-\frac{n(-\xi/\Delta)+n((\xi+U)/\Delta)+4n_2(-\xi/\Delta,(\xi+U)/\Delta)}{\Delta}
    \qquad\mathrm{for}\quad -U<\xi<0,\\
    N(\xi) &=\int_{\Delta}^{\infty} dE \int_{\Delta}^{\infty} dE' \frac{2\Delta^2}{\sqrt{E^{2}-\Delta^{2}}\sqrt{E'^{2}-\Delta^{2}}}
    \left[\frac{1}{(E+E')\left(\xi+U-E\right)\left(\xi+U-E'\right)}-\frac{2}{(2\xi+U)(\xi+U-E)(\xi+U-E')}\right]\nonumber\\
    &=2\left(\frac{n(|\xi+U|/\Delta)}{\Delta}-\frac{\arccos^2(|\xi+U|/\Delta)}{(\xi+U/2)(1-(|\xi+U|/\Delta)^2)}\right)
    \qquad\mathrm{for}\quad\xi<-U.
\end{align}
\end{subequations}
\end{widetext}
Here, $n(x)$ is defined by Eq.~\eqref{funct_n} and
\begin{equation}\label{func_n2}
\begin{split}
    n_2(x,y)&=\int_1^{\infty}\frac{du}{\sqrt{u^2-1}}
    \int_1^{\infty}\frac{dv}{\sqrt{v^2-1}}
    \frac{1}{(u+v)(u+x)(u+y)}\\
    &=\int_1^{\infty}du \frac{\ln(u+\sqrt{u^2-1})}{(u^2-1)(u+x)(u+y)}.
\end{split}
\end{equation}
Again, the appropriate analytic continuation of $\arccos x$ for
$x>1$ is understood (see discussion below Eq.~\eqref{funct_j}).
While neither of the integral definitions of $n(x)$
(Eq.~\eqref{funct_n}) or $n_2(x,y)$ (Eq.~\eqref{func_n2}) seems to
lead to an explicit formula we still could, however, use the above
equations for indirect evaluation of $n(x)$. In particular, since
the first equation \eqref{finiteU} is valid for $\xi>0$ for any
value of $U$ including both the limits $U\to 0$ and $U\to\infty$
we can use the former limit for the evaluation of the latter one.
The noninteracting case $U=0$ is described by Eq.~\eqref{Ic0} and,
thus, we come to the simple result \eqref{funct_j} (in the limit
$\beta\to\infty$). The function $n_2(x,y)$ can be easily
calculated by numerical evaluation of the integral in
Eq.~\eqref{func_n2} for $x,y>-1$.

\end{document}